\documentclass{jfm}
\usepackage{graphicx,color}
\usepackage{float}
\usepackage{epstopdf, epsfig}
\usepackage{caption,subcaption}
\usepackage[intlimits]{amsmath}
\usepackage{lscape}
\usepackage{rotating}
\usepackage{amssymb, amsmath}
\usepackage{amsmath}
\usepackage{xcolor}

\newcommand{\IM}{\mathcal{M}}

\newcommand{\Koop}{\mathcal{K}}
\newcommand{\Kapp}{\mathbf{K}}

\shorttitle{Koopman and NODE in Invariant Manifolds}
\shortauthor{C. R. Constante-Amores  and M. D. Graham}

\title{
Data-driven state-space and Koopman operator models of coherent state dynamics on invariant manifolds  
}

\author{C. Ricardo Constante-Amores, Michael D. Graham \corresp{\email{mdgraham@wisc.edu}} 
}
  
\affiliation{
Department of Chemical and Biological Engineering, University of Wisconsin-Madison, Madison WI
53706, USA
}

\begin{document}

\maketitle
The accurate simulation of complex dynamics in fluid flows demands a substantial number of degrees of freedom, i.e.~ a high-dimensional state space. Nevertheless, the swift attenuation of small-scale perturbations due to viscous diffusion permits in principle the representation of these flows using a significantly reduced dimensionality. Over time, the  dynamics of such flows evolve towards a finite-dimensional invariant manifold.
Using only data from direct numerical simulations, in the present work we identify the manifold and determine evolution equations for the dynamics on it.
We use an advanced autoencoder framework to  automatically estimate the intrinsic dimension of the manifold  and provide an orthogonal coordinate system. 
Then, we learn the dynamics by determining an equation on the manifold by using both a function space approach (approximating the Koopman operator) and a state space approach (approximating the vector field on the manifold). 
We apply this method to exact coherent states 
for Kolmogorov flow and minimal flow unit pipe flow. Fully resolved simulations for these cases require $\mathcal{O}(10^3)$ and $\mathcal{O}(10^5)$ degrees of freedom respectively, and we build models with two or three degrees of freedom that faithfully capture the dynamics of these flows. For these examples,  both the state space and function space time evaluations provide highly accurate predictions of the long-time dynamics in manifold coordinates. 
\begin{abstract}

\end{abstract}

\section{Introduction}

The Navier-Stokes equations (NSE) are dissipative partial differential equations (PDE) that describe the motion of fluid flows.
When they have complex dynamics, 
their description  requires a large number of degrees of freedom -a high state space dimension- to accurately resolve their dynamics. 
However, due to the fast damping of small scales by viscous diffusion, the long-time dynamics relax to a finite-dimensional surface in state space, an {\it invariant manifold} $\IM$ of embedding dimension $d_{\IM}$ \citep{Temam,Foias,Zelik}. 
The long-time dynamics on $\mathcal{M}$ follow a set of ordinary differential equations in $d_\mathcal{M}$ dimensions;
since $\mathcal{M}$ is invariant under the dynamics, the vector field defined on $\mathcal{M}$ remains tangential to $\mathcal{M}$. 
Classical data-driven methods for dimension reduction, such as Proper Orthogonal Decomposition (POD), approximate this manifold as a flat surface, but for complex flows, this linear approximation is severely limited \citep{holmes_2012}. Deep neural networks have been used to discover the \textcolor{black}{global} invariant manifold coordinates for complex chaotic systems such as the Kuramoto–Sivashinsky equation, channel flow, Kolmogorov flow or turbulent planar Couette flow  \citep{Milano,Page_2021,alec_pre,carlos_prf,irmae,linot_couette}.
\textcolor{black}{ \cite{Floryan} recently introduced an approach in which the global manifold is split into local charts  to  identify the intrinsic dimensionality of the manifold (i.e., minimal-dimensional representation of a manifold often necessitates multiple local  charts). This approach is natural for dealing with discrete symmetries, as illustrated in \cite{carlos_symmetry}. In the present work, we only consider global coordinates in the embedding dimension of the manifold.
}

Turbulent flows exhibit patterns that persist in space and time, often called  coherent structures \citep{waleffe_2001,Kawahara,Graham_ecs}.
In some cases, nonturbulent exact solutions to the NSE exist that closely resemble these structures; these have been referred to as {\it exact coherent structures }`ECS'.
There are several ECS types: steady  or equilibrium solutions, periodic orbits, travelling waves, and relative periodic orbits. 
The dynamical point of view of turbulence describes turbulence as a 
state-space populated
with simple invariant solutions  whose 
stable and unstable manifolds form a framework that guides the trajectory of turbulent flow as it transitions between the neighborhoods of different solutions.
Thus, these simple invariant solutions can be used to reproduce statistical quantities of the spatio-temporally-chaotic systems. This idea has driven great scientific interest in finding ECS for  Couette, pipe, and Kolmogorov flows \citep{nagata_1990,wedin_kerswell_2004,waleffe_2001, Graham,page_kerswell_2020}.

Our aim in the present work is to apply data-driven modeling methods for time evolution of exact coherent states on the invariant manifolds where they lie.
Consider first full-state data $\textbf{x}$ that live in an ambient space $\mathbb{R}^N$, where $d\textbf{x}/dt=\textbf{f}(\textbf{x})$ governs the evolution of this state over time. When $\textbf{x}$ is mapped into the coordinates of an invariant  manifold, denoted $\textbf{h} \in \mathbb{R}^{d_\mathcal{M}}$, a corresponding evolution  equation in these coordinates can be formulated: $d\textbf{h}/dt=\textbf{g}(\textbf{h})$. To learn this equation of evolution either a state space or function space approach can be applied; we introduce these here and provide further details in Section \ref{sec:framework}.
 The learning goal of the state-space modeling focuses on finding an accurate representation of 
$\textbf{g}$. A popular method for low-dimensional systems when data $d{\bf h}/dt$ is available is the `Sparse Identification of Nonlinear Dynamics (SINDy)' \citep{sindy}. SINDy uses sparse regression on a dictionary of terms representing the vector field, and has been widely applied to systems where these have simple structure. A more general framework, known as `neural ODEs' (NODE) \citep{chen2019neural},  represents the vector field $\bf g$ as a neural network and does not require data on time derivatives. It has been applied to complex chaotic systems \cite{alec_chaos,linot_couette} and will be used here.
Function space modeling is based on the Koopman operator, a linear infinite-dimensional operator that evolves observables of the state space forward in time \citep{koopman1931,Lasota}.
To approximate the Koopman operator, a leading method is the Dynamic Mode Decomposition (DMD) which considers the state as the only observable, making it essentially a linear state space model \citep{schmid_2010,Rowley}. Other methods have been proposed to lift the state into a higher-dimensional feature space, such as the
Extended DMD (EDMD) that uses a dictionary of observables where the dictionary is chosen to be Hermite or Legendre polynomial functions of the state \citep{eDMD}.
However, without knowledge of the underlying dynamics, it is difficult to choose a good set of dictionary elements, and data-driven approaches have emerged for learning Koopman embeddings \citep{lusch2018deep, Kaiser_2021}. One of these approaches is  EDMD with dictionary learning (EDMD-DL), in which neural networks are trained as dictionaries to map the state to a set of observables, which are evolved forward with a linear operator \citep{Qianxiao,edmd_dl_ad}. 
\textcolor{black}{In this work  we are applying the Koopman operator theory to the inertial manifold theory. \citet{Nakao} and \citet{mezic2020koopman} showed that inertial manifolds correspond to joint zero level sets of Koopman eigenfunctions.}
\textcolor{black}{We 
also note
that the Koopman theory has been applied to systems with continuous spectrum, see \citet{Arbabi,resdmd}.
}

\textcolor{black}{In this work,  we address data-driven modeling from a dynamical system perspective. We show that both state-space and function-space approaches are highly effective when coupled with dimension reduction 
for exact coherent states of the NSE for Kolmogorov flow and pipe flow.}
The rest of this article is organised as follows: Section 2 presents the framework of our methodology. Section 3 provides a discussion of the results, and concluding remarks are summarised in Section 4.

\section{Framework}\label{sec:framework}

\begin{figure}
\begin{center} 
\includegraphics[width=1\linewidth]{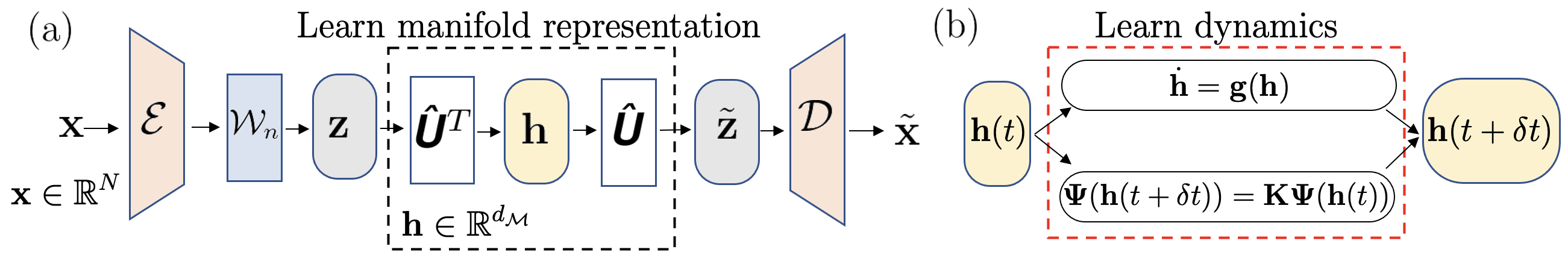}
\end{center} 
\caption{\label{Framework} (a) Representation of the framework for identifying $d_\mathcal{M}$ and $\mathcal{M}$  coordinates. (b) Forecasting on the manifold coordinates   using either neural ODE or Koopman.
} 
\end{figure}

Here we describe the framework to identify the intrinsic manifold dimension $d_\mathcal{M}$, the 
mapping between $\mathcal{M}$  and the full state spaces, and learn the time-evolution model for the dynamics in $\mathcal{M}$. See figure \ref{Framework} for a schematic representation.
We assume that our data comes in the form of  snapshots, each representing the full state from a long time series obtained through  a fully-resolved direct numerical simulation. With the full space,
we use a recently-developed IRMAE-WD (implicit rank-minimizing autoencoder-weight decay) autoencoder architecture \citep{irmae} to identify the mapping into the manifold coordinates
${\bf h}=\boldsymbol{ \chi}(\tilde{\bf x})$, along with a mapping back $\tilde{\bf x}=\tilde{ \boldsymbol{\chi}}({\bf h})$, so these functions can in principle reconstruct the state (i.e., ${\bf x} =\tilde{\bf x}$).
The autoencoder is formed by  standard nonlinear encoder and decoder networks with $n$ additional linear layers with weight matrices $W_1, W_2,\ldots, W_n$ (of size $d_z \times d_z$) signified by $\mathcal{W}_n$ in Figure \ref{Framework} between them \textcolor{black}{ (see Table 1 for the architectures of the networks used in this work)}. The encoder finds a compact representation ${\bf z} \in \mathbb{R}^{d_z}$, and the decoder performs the inverse operation. 
The additional linear layers promote minimization of  the rank of the data covariance in the latent representation, precisely aligning  with the dimension of the underlying manifold. 
Post-training, a singular value decomposition (SVD) is applied to the covariance matrix of the latent data matrix  $\textbf{z}$
yielding matrices of singular vectors $\mathsfbi{U}$, and  singular values $\mathsfbi{S}$. 
Then, we can 
\textcolor{black}{ recast $\textbf{z}$ in the orthogonal basis given by the columns of $\mathsfbi{U}$ by defining $\textbf{h}^\times=\mathsfbi{U}^T \textbf{z} \in\mathbb{R}^{d_z}$}, 
in which each coordinate of $\textbf{h}^\times$ is ordered by contribution. 
This framework reveals the manifold dimension $d_\IM$ as the number of significant singular values, indicating that a coordinate representation exists in which the data spans $d_\mathcal{M}$ directions.
Thus, the encoded data avoids spanning directions  associated with nearly zero singular values (i.e., $\mathsfbi{U} \mathsfbi{U}^T {\bf z} \approx  \mathsfbi{\hat{U}} \mathsfbi{\hat{U}}^T {\bf z} $, where $\mathsfbi{\hat{U}}$ are the singular vectors truncated corresponding to singular values that are not nearly zero). Leveraging this insight, we  extract a minimal, orthogonal coordinate system by 
\textcolor{black}{ projecting  $\textbf{z}$  onto the range  of  $\mathsfbi{\hat{U}}$},
resulting in a minimal representation ${\bf h} =\mathsfbi{\hat{U}}^T {\bf z}  \in \mathbb{R}^{d_{\mathcal{M}}}$.
\textcolor{black}{In summary, $\chi$ is defined by $\chi({\bf x}) = \mathsfbi{\hat{U}}^T (\mathcal{W}_n(\mathcal{E}({\bf x})))$, while $\tilde{\chi}({\bf h}) = \mathcal{D}(\mathsfbi{\hat{U}}{\bf h})$}.

In the neural ODE framework for modeling state-space time evolution, we represent the vector field $\bf{g}$ on the manifold as a neural network with weights $\theta_f$ \citep{alec_chaos,linot_couette}. For a given $\bf{g}$, we can time-integrate the dynamical system between $t$ and $t+\delta t$ to yield a prediction $\tilde{\bf{h}}(t+\delta t)$: i.e.,
$\tilde{{\bf h}}(t+\delta t)={\bf h}(t)+\int_{t}^{t+\delta t}{\bf g}({\bf h}(t');\theta_f) dt' $.  Given data for ${\bf h}(t)$ and ${\bf h}(t +\delta t)$ for a long time series we can train $\bf{g}$  to minimize the $L_2$ difference between the prediction $\tilde{{\bf h}}(t+\delta t)$   and the known data  ${\bf h}(t+\delta t)$.
We use automatic differentiation to determine the derivatives of $\bf{g}$ with respect to  $\theta_f$. 
\textcolor{black}{ In some applications of NODE to complex dynamical systems, it has been found useful to add a stabilization term to prevent drift away from the attractor \citep{alec_node}. We did not find this to be necessary here.}

The function-space approach to time evolution is based on the  infinite-dimensional linear Koopman operator $\Koop_{\delta t}$, which describes the evolution of an arbitrary observable $G({\bf h})$ from time $t$ to time $t+\delta t$: $G({\bf h}(t+\delta t))=\Koop_{\delta t}G({\bf h}(t))$ \citep{koopman1931,Lasota, Mezic_2023}. 
The tradeoff for gaining linearity is that $\Koop_{\delta t}$ is also infinite-dimensional, requiring for implementation some finite-dimensional truncation of the space of observables. Here we use a variant of the `extended dynamic mode decomposition-dictionary learning' approach
which performs time-integration of the linear system governing the evolution in the space of observables
\citep{Qianxiao,edmd_dl_ad}.
Given a vector of observables $\boldsymbol{\Psi}({\bf h}(t))$, now there is a matrix-valued approximate Koopman operator $\Kapp$ such that the evolution of observables is approximated by $\boldsymbol{\Psi}({\bf h}(t+\delta t))=\Kapp\boldsymbol{\Psi}({\bf h}(t))$. 
Given a matrix of observables, whose columns are the vector of observables at different times,
$\boldsymbol{\psi}(t)=\left [ \boldsymbol{\Psi}({\bf h}(t_1)),~ \boldsymbol{\Psi}({\bf h}(t_2)) \ldots \right ]$ and its corresponding matrix at  $ t + \delta t$,
$\boldsymbol{\psi}(t+\delta t)=\left [ \boldsymbol{\Psi}({\bf h}(t_1 +\delta t)),~  \boldsymbol{\Psi}({\bf h}(t_2 +\delta t)) \ldots \right ]$, 
the approximate matrix-valued Koopman operator is defined as the least-squares solution 
$\Kapp= \boldsymbol{\psi}({t + \delta t }) \boldsymbol{\psi}(t)^\dagger $, 
where $\dagger$ superscript denotes the  Moore-Penrose pseudoinverse.
We aim to minimise 
$\mathcal{L}({\bf h },\theta) = ||  \boldsymbol{\psi}{({\bf h }, t + \delta t;\theta )(\mathcal{I}- (\boldsymbol{\psi}({\bf h },t;\theta)^\dagger  \boldsymbol{\psi}({\bf h }, t;\theta))} ||_F$,
where $\mathcal{I}$ and $\theta$ stand for the identity matrix and the weights of the neural networks, respectively. Due to
advancements in automatic differentiation, we can now compute the gradient of $\partial\mathcal{L}/\partial \theta$ directly, enabling us to
find $\Kapp$ and the set of observables $\boldsymbol{\Psi}({\bf h})$ simultaneously using the Adam optimizer \citep{Adam}.
For more details, we refer  to \cite{edmd_dl_ad}, \textcolor{black}{in which this Koopman methodology is extensively explained.}

\begin{table}
\captionsetup{justification=raggedright}
\caption{Neural networks architectures. Between the $\mathcal{E}$ and $\mathcal{D}$, there are  $n$ sequential linear layers  \textcolor{black}{$\mathcal{W}_i$} of shape \textcolor{black}{$d_z \times d_z$}  (i.e., $n=4$ and $d_z=10$). 
}
\centering
\begin{tabular}{ll*{6}{c}r}
Case&Function & Shape & Activation  & Learning Rate \\
\hline
{\it Kolmogorov Flow} & $\boldsymbol{\chi}$		& 		1024/5000/1000/$d_z$ \quad           & sig/sig/lin         & $[10^{-3},10^{-4}]$ \\
 & $\check{\boldsymbol{\chi}}$		& $d_z$/1000/5000/1024 \quad           & sig/sig/lin         & $[10^{-3},10^{-4}]$ \\
 & $\boldsymbol{\Psi }$	    &  $ d_\mathcal{M}$/100/100/$\Psi+ d_\mathcal{M}$  \quad  & elu/elu/elu/elu & $[10^{-3},10^{-4}]$\\
  & $\boldsymbol{g}$	    &  $ d_\mathcal{M}$/200/200/$d_\mathcal{M}$ \quad  & sig/sig/lin & $[10^{-3},10^{-4}]$

 \\
{\it Pipe Flow} & $\boldsymbol{\chi}$		& 		508/2000/1000/$d_z$ \quad           & sig/sig/lin         & $[10^{-3},10^{-4}]$ \\
 & $\check{\boldsymbol{\chi}}$		& $d_z$\textcolor{black}{/1000/2000/}508 \quad           & sig/sig/lin         & $[10^{-3},10^{-4}]$ \\
 & $\boldsymbol{\Psi }$	    &  $ d_\mathcal{M}$/100/100/100/$\Psi+ d_\mathcal{M}$ \quad  & elu/elu/elu/elu & $[10^{-3},10^{-4}]$ \\
   & $\boldsymbol{g}$	    &  $d_\mathcal{M}$/200/200/$d_\mathcal{M}$ \quad  & sig/sig/lin & $[10^{-3},10^{-4}]$

\label{Table_Dist_kolmog_large}
\end{tabular}
\end{table}

\section{Results}

\subsection{Kolmogorov flow}

We consider monochromatically forced, two-dimensional turbulence in a doubly-periodic domain (‘Kolmogorov’ flow), for which the governing equations are solved in a domain of size $(x,y)=[0,2\upi]\times [0,2\upi] $. The governing equations are 
\begin{equation}
\begin{aligned}
\nabla \cdot \textbf{u} &= 0, ~~~\
\frac{\partial \textbf{u}}{\partial t} + \textbf{u} \cdot \nabla \textbf{u} + \nabla p &= \frac{1}{\Rey}~ \nabla^2 \textbf{u} + \sin(\mathit{n}_fy) \textcolor{black}{\textbf{e}_x},
\end{aligned}
\end{equation}
where, $\Rey$ and $\mathit{n}_f$ are the Reynolds number  $\Rey=\sqrt{\chi_f}(L_y/2\upi)^{3/2}/\nu$ (here $\chi_f$, $L_y$ and $\nu$ stand for the forcing amplitude, the height of the computational domain and kinematic viscosity, respectively), and the forcing wavelength, respectively. We assume a forcing wavelength $\mathit{n}_f = 2$, as done previously by \citet{carlos_prf}. 
\textcolor{black}{ The dissipation rate and power input for this system are given, respectively, by ($\it{D}= \left \langle | \nabla \textbf{u}^2|\right \rangle_V /\Rey$), and  ($\it{I}=  \left \langle  u \sin (n_fy) \right \rangle_V$),} where the volume average is defined as $\langle \rangle_V  =1/(4\upi^2)\int_0^{2\upi}\int_0^{2\upi}dxdy$.

\textcolor{black}{We consider cases with $\Rey=10$ and $\Rey=12$, where a stable traveling wave (TW) and a stable relative periodic orbit (RPO) exist, respectively.} 
Data was generated using the vorticity representation with $\Delta t =0.005$ on a grid of $[N_x,N_y]=[32,32]$ (e.g., $\omega \in \mathbb{R}^{1024}$) following the pseudospectral scheme described by \citet{chandler_kerswell_2013}. Simulations were initialized from random divergence-free initial conditions, and  evolved forward in time  to  $10^5$  time units. We drop the early transient dynamics and select $10^4$ snapshots of the flow field, separated by $\delta t=5$ and $\delta t=0.5$ time units for $\Rey=10$ and $\Rey=12$, respectively.  We do an $80\%/20\%$ split for training and testing respectively. The neural network training uses only the training data, and all comparisons use test data unless otherwise specified. \textcolor{black}{Finally, we note that Kolmogorov flow has a continuous translation symmetry as well as several discrete symmetries. While we have not done so here, it is possible to exploit these to further improve the effectiveness of data-driven models  \citep{alec_pre,linot_couette,carlos_symmetry}.}

\subsubsection{Travelling wave (TW) at $\Rey=10$}

\begin{figure}
\begin{center} 
\includegraphics[width=1\linewidth]{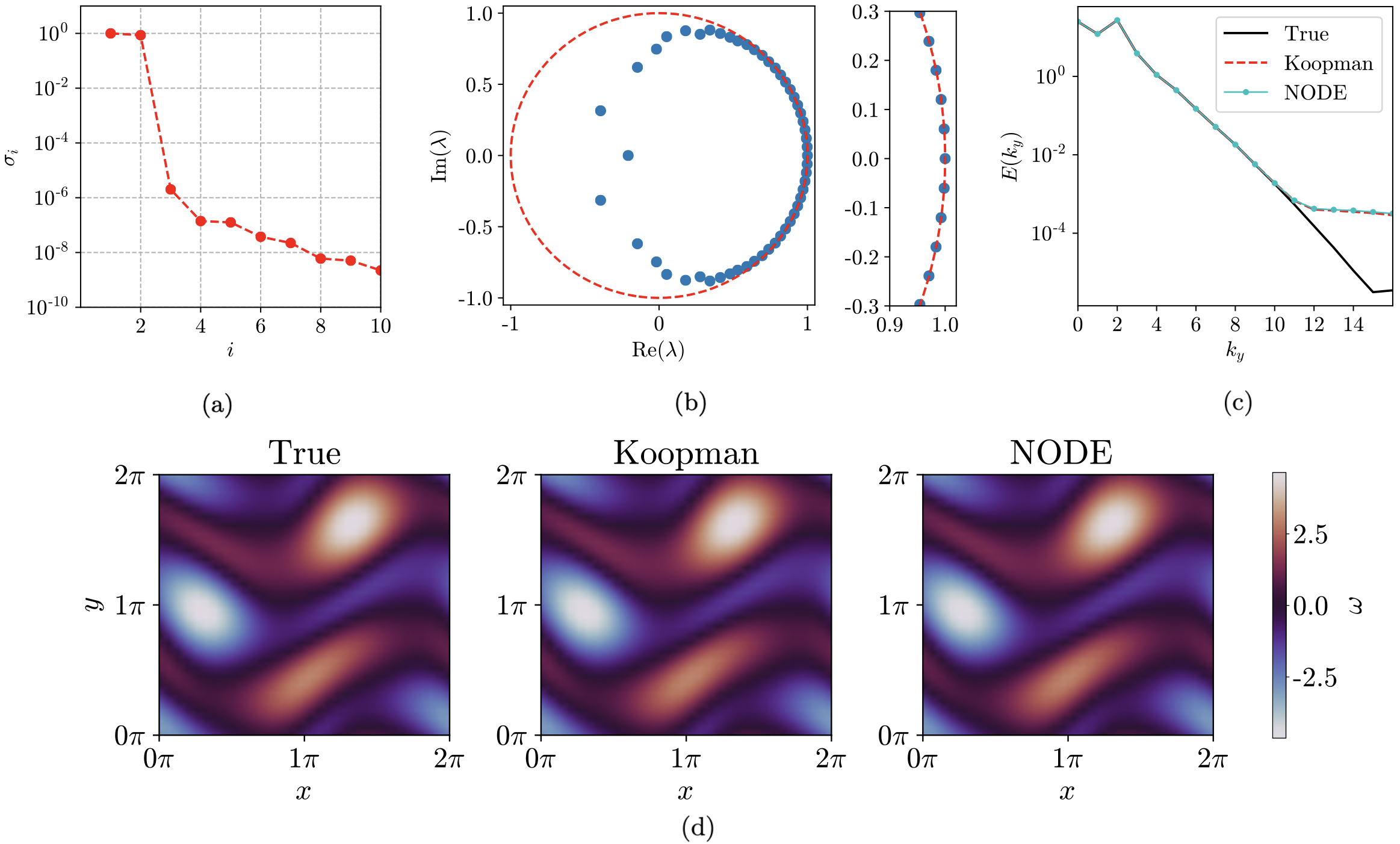}
\end{center} 
\caption{\label{TW_figure}
Kolmogorov flow with $\Rey=10$:  traveling wave regime.
(a) Identification of  $\mathcal{M}$: normalised singular values of the latent space data with a drop at $d_\mathcal{M}=2$.
 (b) Eigenvalues of the approximate Koopman operator (the right panel shows a magnified view of the  eigenvalues).  (c) Energy spectrum.
(d) Snapshots of the vorticity field for the ground truth and models  after 5000 time units evolved with the same initial condition. 
} 
\end{figure}

Figure \textcolor{black}{\ref{TW_figure}}a shows the singular values $\sigma_i$ resulting from performing the SVD on the covariance matrix of the latent data matrix $\textbf{z}$ generated with IRMAE-WD for a TW with period $T=161.45$. 
The singular values for $i>2$ drop to $\approx10^{-6}$ indicating that $d_\IM=2$. This is the right embedding dimension for a TW, as the embedding dimension for a limit \textcolor{black}{cycle} is two.

Once, we have found the mapping to the manifold coordinates, we apply both the function and state approaches, evolving the same initial condition forward in time out to 5000 time units (e.g., $30.96$ periods). Figure \ref{TW_figure}b shows the eigenvalues, $\lambda_k$, of the approximated Koopman operator (with $50$ dictionary elements), and some of them are located on the unit circle, i.e.,  $|\lambda_k|=1$, implying that the dynamics will not decay. Any contributions from an eigenfunction with $|\lambda_k| < 1$ decay as $t \rightarrow \infty $, and the fact that the dynamics live on an attractor, prohibits any $\lambda_k$ from having $|\lambda_k| > 1$. 
\textcolor{black}{As shown in the magnified view of figure \ref{TW_figure}b, the eigenvalues of the Koopman operator yields eigenvalues that are multiples to its fundamental frequency (in agreement with \citet{mezic2005spectral}). 
We also observe eigenvalues within the unit circle, a phenomenon that
seems counterintuitive given that the dynamics are on-attractor, after transient have decayed; thus,  all eigenvalues should inherently be on the unit circle. We attribute this characteristic to be an artefact of  EDMD-DL as it has also been observed  by \citet{edmd_dl_ad}. Several approaches have been proposed   to address this issue by explicitly constraining the eigenvalues to unity, as explored by  \citet{piDMD} and \citet{mpEDMD}.}

Figure \ref{TW_figure}c displays the spatial energy spectrum of the dynamics from the data and both data-driven time-evolution methods, demonstrating that both approaches capture faithfully the largest scales (with a peak in $k_y=2$ corresponding to the forcing) up to a drop of three orders of magnitude that reaffirms the accuracy of both the NODE and the Koopman predictions. 
However, the models cannot capture the highest wavenumbers (smallest scales), $k\geq 11$, of the system (e.g., the discrepancies observed in the highest wavenumbers are a consequence of the dimension reduction performed through IRMAE-WD). To further evaluate the accuracy of our method to capture the long-time dynamics, we compare the average rate of dissipation and input between the true data and the models.
For the true data,  $\it{D}=I=0.267350$, and both the NODE and Koopman approaches reproduce these with relative errors of $<10^{-4}$.
Finally, figure \ref{TW_figure}d shows vorticity field snapshots of the system data and predictions after 5000 time units.
Both the Koopman  and NODE approaches are capable of capturing accurately the  dynamics of the system in $\mathcal{M}$, enabling accurate predictions over extended time periods.

\subsubsection{Relative Periodic Orbit (RPO) at $\Rey=12$}

\begin{figure}
\begin{center} 
\includegraphics[width=0.85\linewidth]{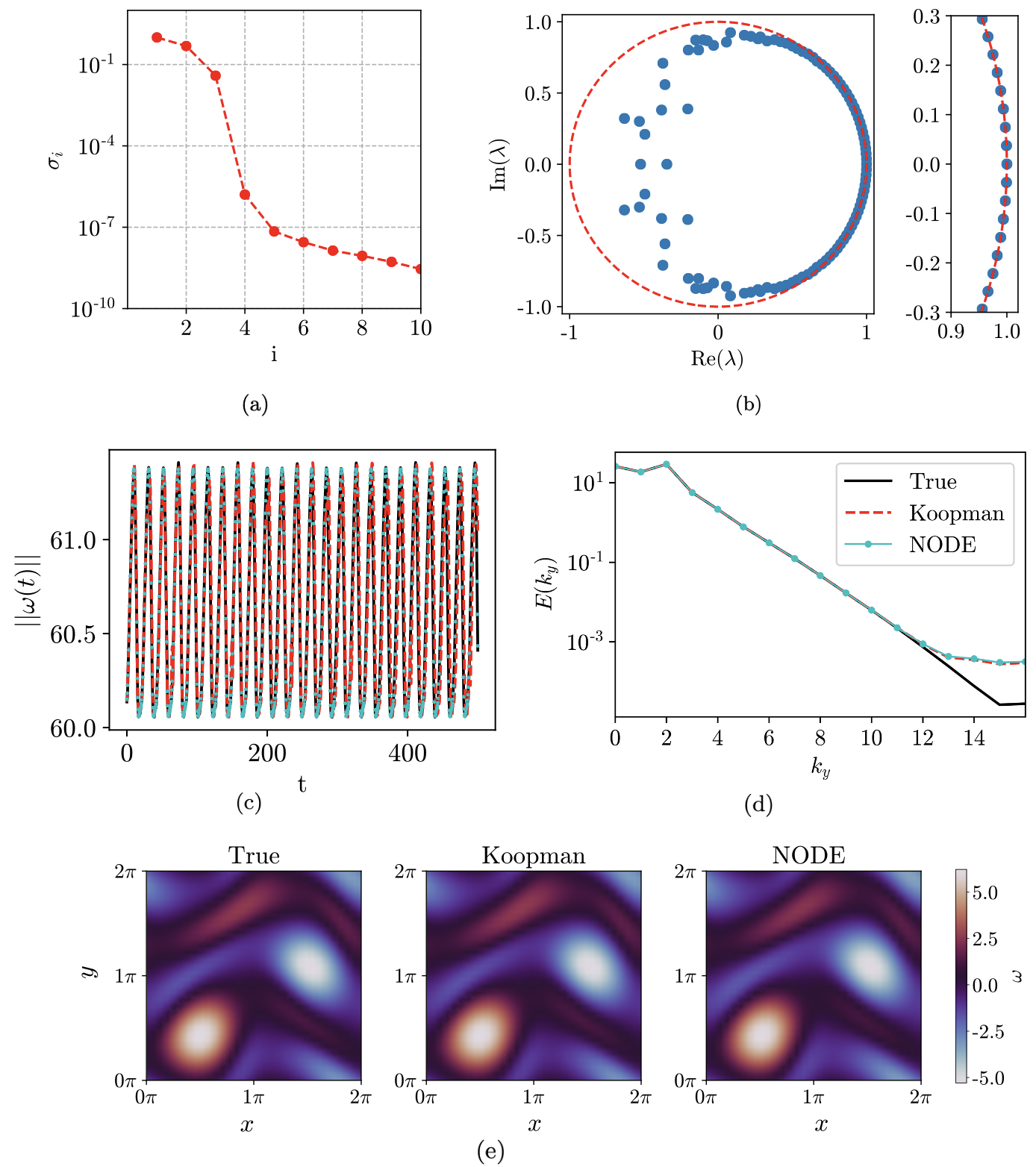}
\end{center} 
\caption{\label{PO_figure}
Kolmogorov flow with $\Rey=12$: relative periodic orbit regime.
(a) Identification of  $\mathcal{M}$:
normalised singular values of the latent space data with a drop at $d_\mathcal{M}=3$.
(b) Eigenvalues of the Koopman operator (the right panel shows a magnified view of the  eigenvalues).
(c) Comparison of $||\omega(t)||$.
(d) Energy spectrum.
(e) Snapshots of $\omega$ for the ground truth and models,  after 500 time units evolved with the same initial condition. 
} 
\end{figure}

Next, we consider Kolmogorov flow at $\Rey=12$, where a stable RPO appears (with period $T=21.29$).
Figure \ref{PO_figure}a shows the singular values $\sigma_i$ resulting from the SVD of the covariance latent data matrix. The singular values for  $i \textcolor{black}{>} 3$ drop to $\approx10^{-6}$, suggesting that the intrinsic dimension of the invariant manifold is $d_\mathcal{M}=3$. This  
is the correct embedding dimension for a RPO, as a dimension of 2 corresponds to the phase-aligned reference frame for a periodic orbit, and one dimension corresponds to the phase.

Now we apply the function and state approaches for the dynamics on $\mathcal{M}$, evolving the same initial condition  in time for 500 time units (e.g., $23.48$ periods). Figure \ref{PO_figure}b shows the eigenvalues of the Koopman operator (with $100$  dictionary elements), and identically to the previous case, some of them have unit magnitude $|\lambda_k|=1$; thus, the dynamics  will neither grow nor decay as time evolves.
Figure \textcolor{black}{\ref{PO_figure}}c and \textcolor{black}{\ref{PO_figure}}d displays the norm of the vorticity $||\omega(t)||$ and the energy spectrum, respectively. For the energy spectrum, both models can capture faithfully the largest scales (with a peak in $k_y=2$ corresponding to the forcing) up to a drop of three orders of magnitude that reaffirms the accuracy of both NODE and  Koopman predictions. 

To further evaluate the accuracy of our method to capture the long-time dynamics, we compare the average rate of dissipation and input between the true data and the models.
For the true data, the time averages  $\it{D}=\it{I}=0.29921$, while for the NODE and Koopman approaches, the predictions again differ with relative errors under $10^{-4}$.
Finally, figure \ref{PO_figure}d shows a vorticity snapshot of the system after 500 time units.
Our analysis reveals that both
the Koopman  and NODE approaches are capable of capturing accurately the   dynamics of the system in $\mathcal{M}$, enabling accurate predictions over extended time periods as their snapshots  closely matches the ground truth.

\subsection{RPO in minimal pipe flow}

\begin{figure}
\begin{center} 
\begin{tabular}{c}
\includegraphics[width=0.88\linewidth]{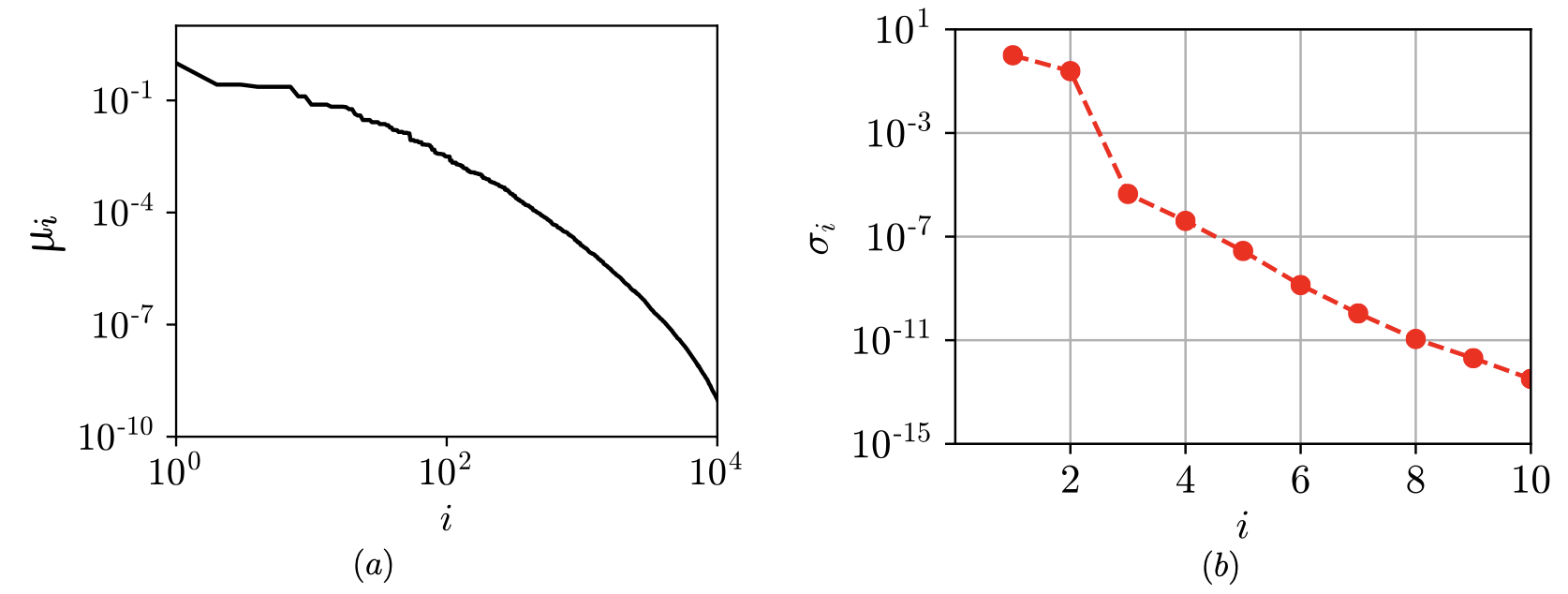}\\
\includegraphics[width=0.89\linewidth]{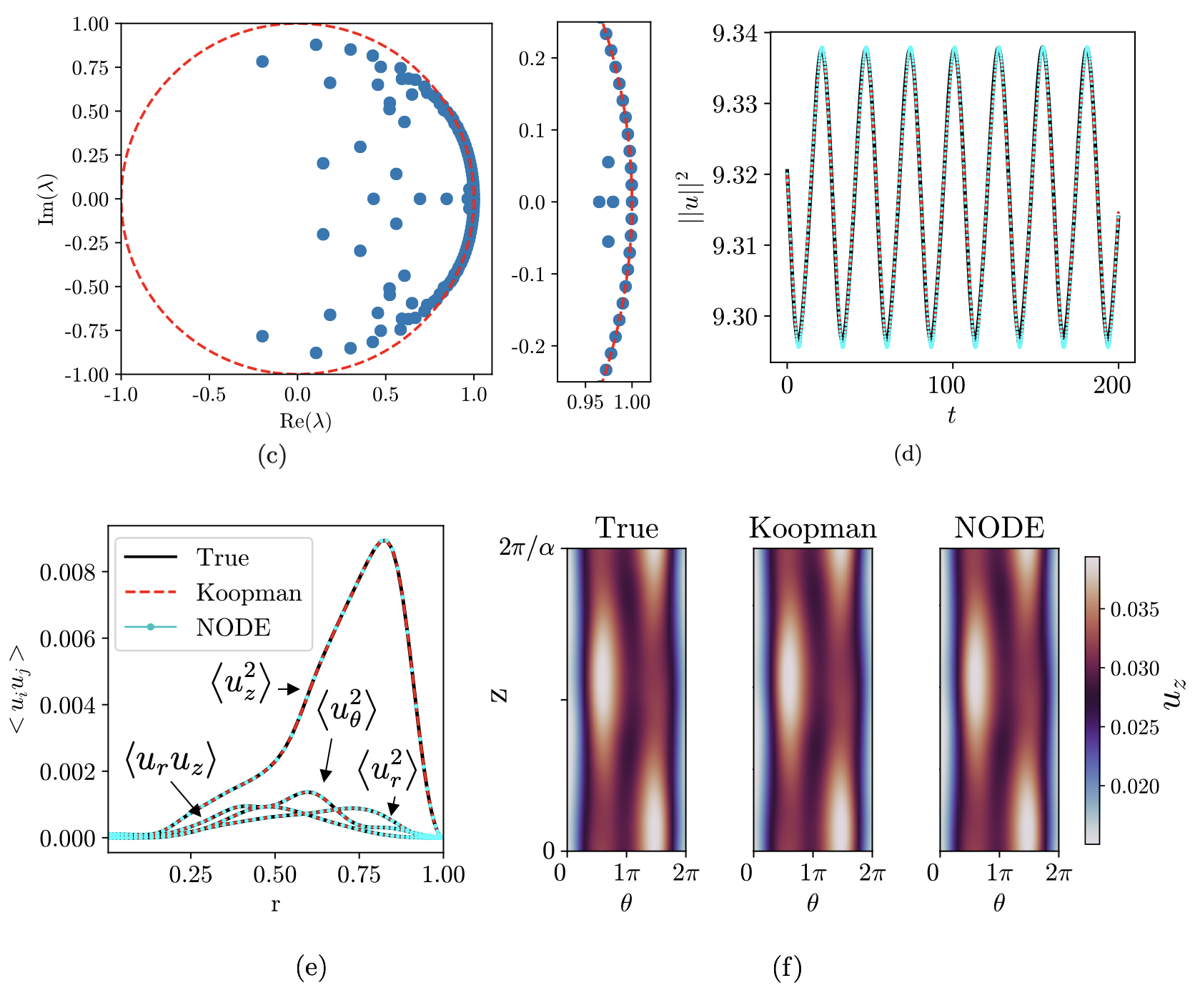}\\
\end{tabular}
\end{center} 
\caption{\label{Pipe_fig}
Pipe flow with $\Rey=2500$: periodic orbit regime.
(a) Eigenvalues of POD modes sorted in descending order.
(b) Identification of the $\mathcal{M}$: normalised singular values of the latent space data with a drop at $d_\mathcal{M}=2$.
(c) Eigenvalues of the Koopman operator (the right panel shows a magnified view of the  eigenvalues).
(d) Comparison of the norms of the velocity field between the models and true data.
(e) Reynolds stresses varying with the radial position from the ground truth, Koopman and NODE predictions, the labels are on the plot.
(f) 
Two-dimensional representation of the dynamics  in a $z-\theta$ plane $(r = 0.496)$ with $u_z$ for the  true  and predicted dynamics at $t=200$. 
} 
\end{figure}

We turn our attention to an RPO with period $T=26.64$ in pipe flow, whose ECS have been found to closely resemble the near-wall quasistreamwise vortices that characterize wall turbulence \citep{willis_avila_2013}.   Pipe flow exhibits inherent periodicity in the \textcolor{black}{azimuthal} direction and maintains a consistent mean streamwise velocity.
Here,  the fixed-flux Reynolds number is $\Rey=DU/\nu=2500$, where length and velocity scales are nondimensionless by the diameter and velocity, respectively. We consider the minimal flow unit in the $m=4$  rotational space (`shift-and-reflect' invariant space); thus, $\Omega=[1/2,2\upi/m,\upi/\alpha]\equiv{(r,\theta,z)\in [0,1/2]\times[0,2\upi/m]\times [0,\upi/\alpha]}$, where $L=\upi/\alpha$ stands for the length of the pipe and $\alpha=1.7$ (as in previous work from  \cite{willis_avila_2013} and \cite{budanur_jfm_2017}). In wall units, this domain is $\Omega^+ \approx [100,160,370]$, which compares well with the minimal flow units for Couette flow and channel flow.

Data was generated with  the pseudo-spectral code Openpipeflow  with $\Delta t=0.01$ on a grid $(N_r,M_\theta,K)=(64,12,18)$, then, following the $3/2$ rule, variables are evaluated on $64\times36\times54$ grid points; thus,
$u\in\mathbb{R}^{124,416}$ 
\citep{willis2017openpipeflow}. We ran simulations forward on time, and stored  $10^3$ time units at intervals of $0.1$ time units. Pipe flow is characterised by  the presence of continuous symmetries, including translation in the streamwise direction and azimuthal rotation about the pipe axis. We phase-align the data for both continuous symmetries using the first Fourier mode method of slices to improve the effectiveness of the dimension reduction process \citep{method_slices}.

To find the manifold dimension and its coordinates, we  first perform  a linear  dimension reduction  from $\mathcal{O}(10^5)$ to $\mathcal{O}(10^2)$ with POD. 
Figure \ref{Pipe_fig}a displays the eigenvalues, $\mu_i$, of POD modes sorted in descending order. \textcolor{black}{
We select the leading 256 modes that capture the $99.99\%$ of the total energy of the system. Most of these modes are characterised by being complex (i.e., 2 degrees of freedom), so projecting onto these modes results in a 508-dimensional POD coefficients.} 
Next, we perform nonlinear dimension reduction in the POD coordinates using IRMAE-WD. Figure \ref{Pipe_fig}b shows the singular values $\sigma_i$ resulting from performing SVD on the covariance matrix of the latent data matrix from the autoencoder. 
The singular values for $i \textcolor{black}{>} 2$ drop to $\approx10^{-6}$, indicating that the dimension of the manifold is $d_\mathcal{M}=2$ which is the correct embedding dimension for a 
\textcolor{black}{ relative} periodic orbit, 
\textcolor{black}{ 
given that we have factored out the continuous symmetries as noted above.} We reiterate that we have elucidated  the precise embedding dimension of the manifold, commencing from an initial state dimension of $\mathcal{O}(10^{5})$.

Having found the mapping to $\mathcal{M}$, we apply both the function and state approaches on $\mathcal{M}$, and evolve the same initial condition forward on time  out to 200 time units (e.g., 7.5 periods). Figure \ref{Pipe_fig}c shows the eigenvalues of the Koopman operator  (with $100$  dictionary elements), and identically to the previous cases, some of them have unit magnitude $|\lambda_k|=1$; thus, the long time dynamics  will not decay.
Figure \ref{Pipe_fig}d displays time evolution of the norm of the velocity field in which the NODE and Koopman predictions can capture the true dynamics (here the norm is defined to be the $L_2$ norm 
\textcolor{black}{$\left \| {\bf u} \right \|_2^2=(1/V)\int {\bf u}\cdot{\bf u}\;d{\bf r}$).}
To further demonstrate that the evolution of the dynamics on $d_\mathcal{M}$  is sufficient to
represent the state in this case, we examine the reconstruction of statistics. In figure \ref{Pipe_fig}e, we show the reconstruction of four components of the Reynolds stress $ \left \langle u_z^2 \right \rangle, \left \langle u_\theta^2 \right \rangle, \left \langle u_r u_z \right \rangle$ and $\left \langle u_r^2 \right \rangle$ (the remaining two components are relatively small). The Reynolds stresses evolved on $d_\mathcal{M}$   closely match  the ground truth.
Lastly,  \ref{Pipe_fig}f displays field snapshots in the $z-\theta$ plane ($r=0.496$) at $t=200$, respectively showing qualitatively that the models can capture perfectly the dynamics of the true system.

\section{Conclusion}

In this study, we have presented a framework that leverages data-driven approximation of the Koopman operator,   and  neural ODE  to construct minimal-dimensional models for exact coherent states to the NSE within manifold coordinates. Our approach integrates an advanced autoencoder-based method to discern the manifold dimension and coordinates describing the dynamics.  Subsequently, we learn the dynamics using both function-space- and state space-based approaches within the invariant manifold coordinates.
We have successfully applied this  framework to construct models for  exact coherent states found in  Kolmogorov flow and minimal flow unit pipe flow. In these situations, performing fully resolved simulations would necessitate a vast number of degrees of freedom. However, through our  methodology, we have effectively reduced the system's complexity from approximately $\mathcal{O}(10^5)$ degrees of freedom to a concise representation comprising fewer than $3$ dimensions, which is capable of faithfully capturing the dynamics of the flow such as the Reynolds stresses for pipe flow, and the average rate of dissipation and power input for Kolmogorov flow.
These results illustrate the capability of nonlinear dimension reduction with autoencoders to identify dimensions and coordinates for invariant manifolds from data in complex flows as well as the capabilities of both state-space and function-space methods for accurately predicting time-evolution of the dynamics on these manifolds.

\textcolor{black}{
By accurately modelling coherent state dynamics with far fewer degrees of freedom than required for  DNS, 
manifold dynamics models like those reported here  open the possibility for dynamical-systems type analyses such as calculation of Floquet multipliers or local Lyapunov exponents in a computationally highly efficient way.
These models also facilitate the application of control strategies with minimal degrees of freedom. An illustrative example is the application of reinforcement learning techniques in controlling planar MFU Couette flow (see \cite{couette_control}).  The linear representation of the dynamics via the Koopman operator facilitates control using well-established techniques, such as Linear Quadratic Regulator (LQR), resulting in efficacious optimal control
for nonlinear systems \citep{Otto}.}

\textcolor{black}{ Finally, we  have also successfully showcased the efficacy of both the function and the state space approaches in handling coherent state dynamics.
When implementing the function space approach, we underscored the critical significance of dimension reduction for the construction of the observable space.
When applying EDMD-DL on the full-state data for pipe flow, the computational challenges escalate significantly when attempting to handle an observable space dimensionality exceeding $\mathcal{O}(10^5)$. Therefore, dimension reduction techniques are pivotal in making 
both the function and the state space approaches
feasible and computationally efficient, ensuring practical applicability to real-world scenarios. } 
\\

This work was supported by  ONR N00014-18-1-2865 (Vannevar Bush Faculty Fellowship).

\bibliographystyle{jfm}
\bibliography{jfm-instructions.bib}

\end{document}